\documentclass[lettersize,journal]{IEEEtran}
\usepackage{amsmath,amsfonts}
\usepackage{algorithm}
\usepackage{array}
\usepackage{textcomp}
\usepackage{stfloats}
\usepackage{url}
\usepackage[colorlinks=true, allcolors=blue]{hyperref}
\usepackage{verbatim}
\usepackage{graphicx}
\usepackage{cite} 
\usepackage{booktabs}
\usepackage{multicol}
\usepackage{graphicx}
\usepackage[export]{adjustbox}
\usepackage{array}
\usepackage{subcaption}
\usepackage{xcolor}
\usepackage{algorithm}
\usepackage{algpseudocode} 
\usepackage{amsmath} 

\newcolumntype{P}[1]{>{\centering\arraybackslash}p{#1}}

\begin{document}
\bstctlcite{IEEEexample:BSTcontrol}
\title{Flexure-FET-Based Receiver with \\ Competitive Binding for Interference Mitigation in Molecular Communication}

\author{Dilara~Aktas,~\IEEEmembership{Graduate Student Member,~IEEE,}
       Ozgur B.~Akan,~\IEEEmembership{Fellow,~IEEE}
     
\thanks{The authors are with the Center for neXt-generation Communications (CXC), Koç University, Turkey. O.B. Akan is also with the Internet of Everything Group, Department of Engineering, University of Cambridge, U.K. (e-mail: dilaraaktas20@ku.edu.tr, oba21@cam.ac.uk).}

\thanks{This work was supported in part by the AXA Research Fund (AXA Chair for Internet of Everything at Ko\c{c} University).}
}

\markboth{}%
{}

\maketitle

\begin{abstract}

Molecular communication (MC), a biologically-inspired technology, enables applications in nanonetworks and the Internet of Everything (IoE), offering unprecedented potential in intra-body systems for applications such as drug delivery, health monitoring, and disease detection. In this paper, we extend our previous research on the Flexure-FET MC receiver by integrating a competitive binding model to enhance its performance in high-interference environments, where multiple species coexist in the reception space. While previous studies have primarily focused on ligand concentration estimation and detection methods, they have not fully addressed the competition between molecular species for receptor binding sites. The competitive binding framework introduced here plays a pivotal role in multitarget environments, closely reflecting real-world biological scenarios. By accounting for the competition between species for receptor binding sites, this framework enhances the understanding of MC dynamics in complex environments. Leveraging this framework enables the fine-tuning of receptor responses, optimizing Flexure-FET MC receiver performance through adjustments in ligand concentrations and receptor affinities. Through comprehensive performance analysis, we demonstrate that integrating competitive binding underscores the importance of accounting for interference in MC systems to improve reliability and accuracy. While the signal-to-noise ratio (SNR) and symbol error probability (SEP) are impacted by factors such as interferer concentration and receptor dynamics, the framework emphasizes the need to manage these interference factors to optimize the system's overall performance. The results show that accounting for interference—through competitive binding—provides a more realistic understanding of system behavior and enables the fine-tuning of the receiver’s response, ultimately ensuring more reliable and accurate detection in environments where multiple species are present.

\end{abstract}


\section{Introduction}

\IEEEPARstart{M}{olecular} communication (MC) is a groundbreaking, biologically-inspired approach that utilizes molecules to encode, transmit, and receive information, mirroring the communication processes found in living organisms \cite{akyildiz2015internet, nakano2013molecular}. This method is essential for communication between cells and bacteria, particularly within the human body, where molecular signals play a key role in physiological functions. Operating at the nanoscale, MC offers a promising alternative in environments where traditional communication methods are either inefficient or impractical. It holds significant potential for applications in nanonetworks and the Internet of Everything (IoE), with transformative possibilities across industries, especially in healthcare \cite{nakano2012molecular, akan2016fundamentals, akan2023internet}.

As nanotechnology advances, MC is increasingly recognized for its role in enabling innovative medical applications such as intrabody communication systems, body-area networks, and real-time health monitoring devices. Moreover, MC plays a crucial role in developing artificial nanonetworks, where nanomachines are interconnected through the Internet of Nano Things (IoNT). This integration facilitates communication between nanoscale systems and larger networks, expanding possibilities for continuous health monitoring, smart drug delivery, disease detection, and the development of artificial organs. These emerging technologies offer reliable, energy-efficient, and secure communication solutions, which are essential for next-generation medical and technological advancements \cite{al2019internet, felicetti2016applications, kuscu2015internet}.

Device modeling for the IoNT, including bio-cyber interface design, is a rapidly developing research area \cite{el2020mixing, chude2016biologically, akyildiz2019microbiome, chude2023crispr, bakhshi2023hybrid}. While much of the work on MC transceivers has focused on biological component-based nanomachines \cite{nakano2014molecular, unluturk2015genetically}, these systems, despite their high biocompatibility, face significant limitations, including insufficient computational capabilities, restricted use to \textit{in vivo} environments, and poor compatibility with larger-scale cyber networks. These limitations hinder the full realization of the IoNT concept, making it difficult to scale these systems for broader applications \cite{unluturk2015genetically, akyildiz2015internet}. 

In contrast, artificial structure-based MC receivers can be utilized in both \textit{in vivo} and  \textit{in vitro} settings \cite{kuscu2019transmitter}. These receivers meet essential functional criteria, such as in situ, continuous, and label-free operation \cite{kuscu2016physical}, and are capable of selectively detecting information ligands, converting these interactions into processable signals.  MC receivers (i.e., MC-Rxs) and biosensors both rely on selective recognition units to measure analyte concentrations, with the integration of various biosensing technologies—electrical, optical, and mechanical—enhancing their versatility and effectiveness for real-world applications.

Building on these advancements, recent progress has also seen the integration of microfluidic systems with artificial MC-Rxs to test and validate their performance. These systems mimic biological environments such as blood vessels, offering controlled flow conditions and facilitating convection-diffusion-based molecular transport \cite{walter2023real}. The incorporation of chemical sensors into these systems has made BioFET-based MC-Rxs highly promising, providing signal amplification, miniaturization, and enhanced selectivity—all of which are essential for their application in real-world scenarios \cite{abdali2024frequency}.

In parallel, recent advancements in nanomaterials, such as carbon nanotubes (CNTs) and graphene, have further enhanced the performance of FET-based biosensors, particularly BioFETs. These materials help BioFET-based MC receivers meet critical requirements for MC systems \cite{kuscu2016physical, kuscu2016modeling}. However, despite their high sensitivity, BioFET-based MC receivers still face limitations, such as their inability to detect neutral ligands because they rely on charged ligands for signal transduction. Moreover, the screening effect caused by high ion concentrations reduces the Debye length, making these systems less effective in certain environments. While aptamers offer a potential solution, their technology is still in development, whereas natural antibodies and enzymes remain more readily available \cite{kuscu2016physical}.

\begin{figure}[!t]
\centering
\includegraphics[width=0.8\columnwidth]{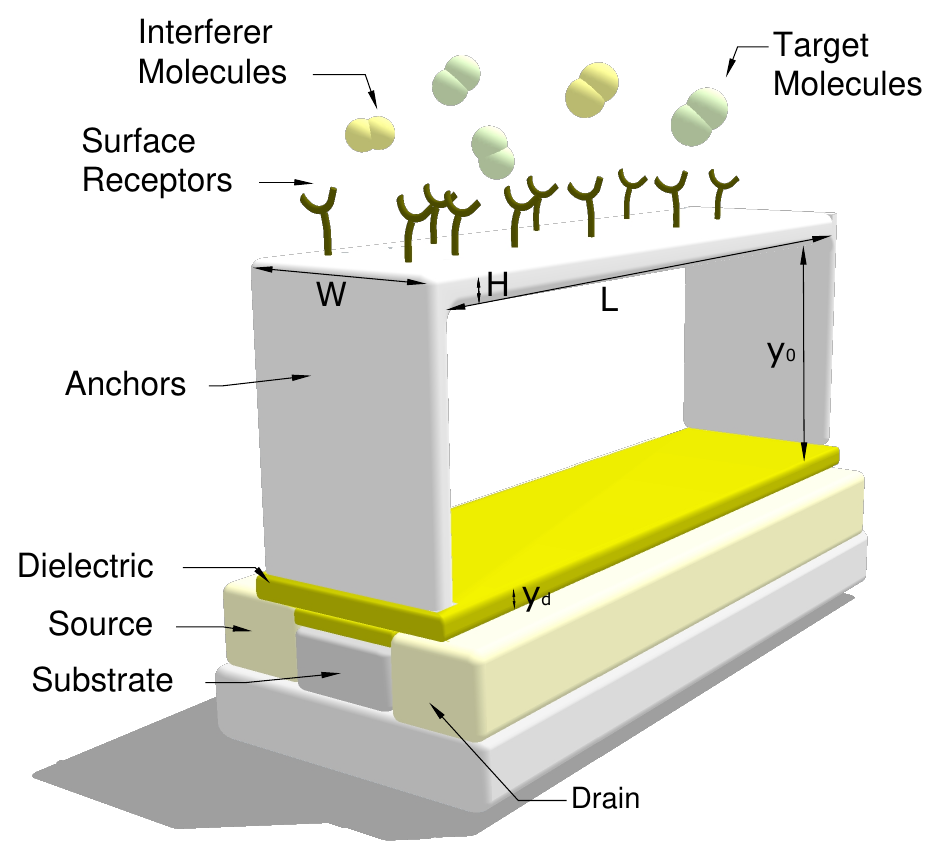}
\caption{The Flexure-FET-based MC receiver, with $W$ denoting the width, $L$ the length, $H$ the microbeam's thickness, $y_d$ the dielectric thickness, and $y_0$ the air gap.}
\label{fig1}
\end{figure}

To address these challenges, we proposed a Flexure FET-based MC receiver \cite{aktas2021mechanical}, as seen in Fig. \ref{fig1}, which utilizes mechanical transduction to detect both charged and neutral molecules, such as proteins and viruses. Building on this concept, we introduced the first practical approach to a WSK-based MC system using an improved Flexure-FET-based MC receiver in \cite{aktas2022weight}. This system leverages the receiver's transduction mechanism, which depends on molecular weight, and we analyze its key performance metrics from a theoretical MC perspective, incorporating biological interference to provide a more realistic simulation.

In this paper, we extend our previous research by integrating a competitive binding model to enhance the performance of the Flexure-FET-based MC receiver in channels with extreme interference. In our earlier work \cite{aktas2022weight}, interference analysis was conducted under simplified conditions, employing equilibrium-based receptor-ligand binding models and two-state stochastic processes. Building upon these foundational insights, we now adopt the competitive binding framework introduced in \cite{blay2020solving}, which fundamentally accounts for molecular interference during receptor-ligand recognition. Unlike traditional models that assume isolated binding events, this approach considers the simultaneous interaction of multiple ligand species competing for the same receptor, leading to competition for binding sites. This approach efficiently solves binding equilibria without the need for initial value estimation, offering a more realistic characterization of receptor occupancy in environments with multiple species, such as target and interferer molecules.

While previous work on MC systems has explored interference from multiple ligands, these models often focus on ligand concentration estimation or detection methods without fully accounting for the competition between different molecular species for receptor binding sites. For example, in \cite{kuscu2019channel}, the focus was on channel sensing techniques using a single receptor type to estimate ligand concentrations, relying on maximum likelihood (ML) and method of moments (MoM) for estimation. Although their work considers interference, it primarily addresses the estimation of ligand concentrations without fully modeling the competition between different molecular species for receptor binding sites, which can significantly impact the accuracy of estimations in interference-rich environments.

In a similar vein, in \cite{kuscu2022detection}, detection methods for MC systems with ligand receptors were explored, focusing on receptor binding statistics. While their models address interference from multiple ligands, they do not account for the competition between molecular species for the same receptor sites, which is crucial for improving detection accuracy in MC channels with extreme interference. Additionally, \cite{bicen2015interference} examined interference in microfluidic MC systems, focusing on the system's capacity under interference from multiple transmitters. Although their work considers interference, it primarily addresses ligand concentration estimation and does not fully capture the competition between molecular species for receptor binding sites, a factor that can significantly affect the accuracy of estimations in MC channels with extreme interference.

In contrast, this paper introduces a novel approach by adapting the competitive binding framework to model interference in MC systems, where multiple molecular species compete for the same receptor binding sites. This model significantly improves the receiver’s ability to operate in environments with interference from various species. By integrating this model into the Flexure-FET MC-Rx,  a more reliable performance can be enabled in real-world molecular communication applications, where interference from multiple species is inevitable.

Furthermore, the competitive binding framework plays a crucial role in multitarget environments, closely mirroring real-world biological scenarios. By leveraging this framework, we can fine-tune receptor responses, thereby optimizing the performance of MC-Rxs in MC systems. Additionally, its ability to detect specific molecules amidst biologically relevant noise highlights the diagnostic potential of competitive binding, which is essential for applications where biological interference can obscure signals.

Furthermore, the framework presented here can be extended to odor-based molecular communication (OMC) receivers and systems \cite{aktas2024odor}, unlocking a wide range of unprecedented applications in the IoE domain, including healthcare, environmental monitoring, food industry, safety, defense, agriculture, and entertainment. The competitive binding model has the potential to enhance the detection of specific odor molecules by accounting for both concentration variations and competition between species with different receptor affinities, improving accuracy in environments with multiple odor compounds. By incorporating this model into practical MC receivers, like Flexure-FET, we can pave the way to OMC systems, revolutionizing non-invasive sensing and real-time detection across various fields.

The rest of the paper is organized as follows: In Section II, we present the Flexure-FET MC receiver architecture, explaining its operating principles and how it handles signal flow and noise sources in MC channels with extreme interference. In Section III, we introduce the modeling approach for the Flexure-FET MC-Rx and explain how the competitive binding framework is integrated into the receiver to address interference caused by competition between ligand species at the receptor level. Section IV then provides a comprehensive performance analysis, focusing on sensitivity, noise, SNR, and SEP, under varying system, receiver, and molecular parameters. Finally, in Section V, concluding remarks are given.

\begin{figure}[!t]
\centering
\includegraphics[width=\linewidth]{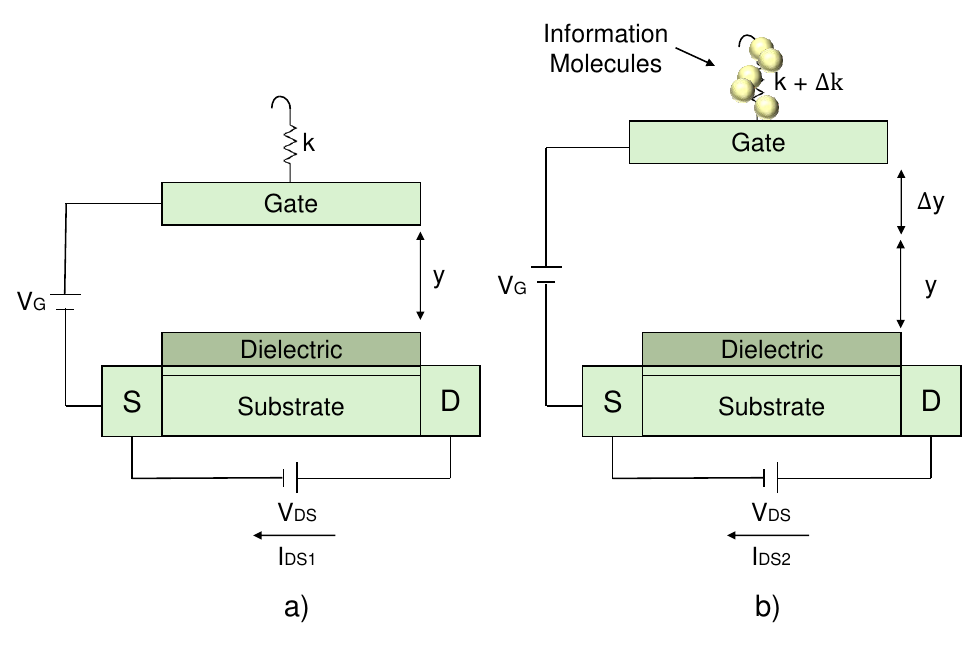}
\caption{Spring-mass system representation of the Flexure-FET: (a) before capture, and (b) after capture \cite{jain2012flexure}.}
\label{fig2}
\end{figure}

\section{Flexure-FET-based MC Receiver Architecture for Interference Mitigation}

The Flexure-FET MC receiver employs mechanical transduction to enhance sensitivity in environments with significant interference, where multiple molecular species coexist in the reception space. This section provides an overview of the receiver's operating principles, including its signal flow and the various noise sources that influence detection accuracy.

\subsection{Flexure-FET Operating Principles and Overview}

Flexure-FET MC receiver employs a mechanical transduction principle to achieve high sensitivity for detecting ligand binding events. Ligands bind to surface receptors covering the upper surface of a suspended gate electrode, which is supported by fixed anchors in a fixed-fixed beam configuration. These binding events modulate the gate stiffness and deflection, altering the air gap, equivalent capacitance, and surface potential. These changes in the mechanical and electrical properties of the gate are transduced into measurable variations in the drain current, as illustrated in Fig.~\ref{fig2}.

The exceptional sensitivity of the Flexure-FET arises from the combination of two electromechanical nonlinearities: \textit{spring softening} and \textit{subthreshold conduction}. Spring softening occurs when the stiffness of the gate decreases nonlinearly with increasing gate voltage (\(V_G\)) until the system reaches the pull-in instability region, a critical point of maximum sensitivity. Subthreshold conduction further amplifies these effects by providing an exponential relationship between surface potential and drain current. Together, these mechanisms allow the Flexure-FET to achieve detection capabilities far beyond conventional nanoscale biosensors. Design optimizations, such as selecting beam materials with a low Young’s modulus (e.g., SU-8) and carefully tuning beam dimensions to lower stiffness while maintaining mechanical flexibility, further enhance the sensitivity and robustness of the device, enabling efficient operation near the critical pull-in instability.

In addition to stiffness-based transduction eliminating the need for reference electrodes, simplifying instrumentation, and enabling the detection of neutral molecules, the Flexure-FET MC receiver offers distinct advantages for interfacing with competitive binding models to account for molecular interference. Its mechanical transduction mechanism directly translates receptor occupancy into measurable changes in stiffness and gate deflection, which can be quantitatively modeled using competitive binding models. This inherent compatibility allows the Flexure-FET to effectively capture the dynamics of interference at the molecular level. Furthermore, the device’s sensitivity amplification near the pull-in instability, coupled with its robustness to ionic screening, makes it particularly suited for interference detection in complex and biologically noisy environments. These characteristics enable precise evaluation of the impact of target and interfering ligands on the system’s performance.

\subsection{Signal Flow and Noise Sources }

In environments with molecular crowding, where multiple molecular species coexist in the reception space, competitive binding influences the interaction between target and interferer molecules, significantly impacting receptor occupancy and the transduced signal. The signal flow diagram in Fig. \ref{signalflow} illustrates these interactions, detailing how various noise sources contribute to fluctuations in the detected output.

\subsubsection{Reception Noise} Molecules in the reception space exhibit random Brownian motion, leading to stochastic binding and unbinding at the receptor sites \cite{kuscu2016physical}. This randomness introduces fluctuations in receptor occupancy, impacting the transduced signal. In the presence of competitive binding, these fluctuations are further amplified as target and interferer molecules compete for receptor sites, increasing variability in molecular recognition. Such noise can hinder accurate concentration detection and reduce the reliability of molecular message decoding \cite{hassibi2005biological}.

\subsubsection{Transduction Noise} In the Flexure-FET MC receiver, the mechanical-to-electrical transduction process is subject to various noise sources that impact signal accuracy. Flicker noise, i.e., $1/f$ noise, originating from charge trapping and defects in the semiconductor material, is particularly dominant at low operating frequencies where MC receivers typically function \cite{kuscu2016physical}. As part of transduction noise, thermomechanical force noise and temperature-induced stiffness fluctuations contribute to random variations in gate position and mechanical stiffness. These sources, along with binding and flicker noise, influence the signal detected by the receiver, affecting its ability to interpret molecular messages accurately. Given the presence of molecular crowding, receptor occupancy is influenced by both signal molecules and biological interference, further amplifying the impact of transduction noise on the overall system response.

\subsubsection{Biological Interference} 

In environments where multiple molecular species coexist in the reception space, receptors in the recognition layer are not exclusively available to target molecules. Instead, interfering molecules with similar binding affinities compete for receptor sites, altering receptor occupancy dynamics and introducing additional noise into the system. This ``biological interference,'' often referred to as background noise, can significantly impact the molecular detection process by reducing receptor availability for target ligands and modulating the transduced signal \cite{kuscu2016physical}. 

Unlike intersymbol interference (ISI) or co-channel interference, which result from signal overlap in time or space, biological interference originates from the inherent competition for receptor sites \cite{kuscu2016physical}. Since interfering molecules may exhibit binding affinities comparable to target ligands, their random binding and unbinding events introduce additional fluctuations in receptor occupancy, further contributing to reception noise \cite{hassibi2005biological}. These effects become more pronounced in environments with high concentrations of interferers, where the binding probabilities of target molecules are significantly reduced and the interferers have higher affinities for the receptors. 

The competitive binding process, influenced by both target and interferer molecules, directly impacts the Flexure-FET MC receiver’s response. The stochastic nature of interferer interactions adds variability to the mechanical transduction process, affecting both the sensitivity and specificity of detection. Therefore, incorporating competitive binding models into the system design is essential for accurately characterizing the influence of biological interference on molecular signal transduction.

\begin{figure}[!t]
\centering
\includegraphics[width=\columnwidth]{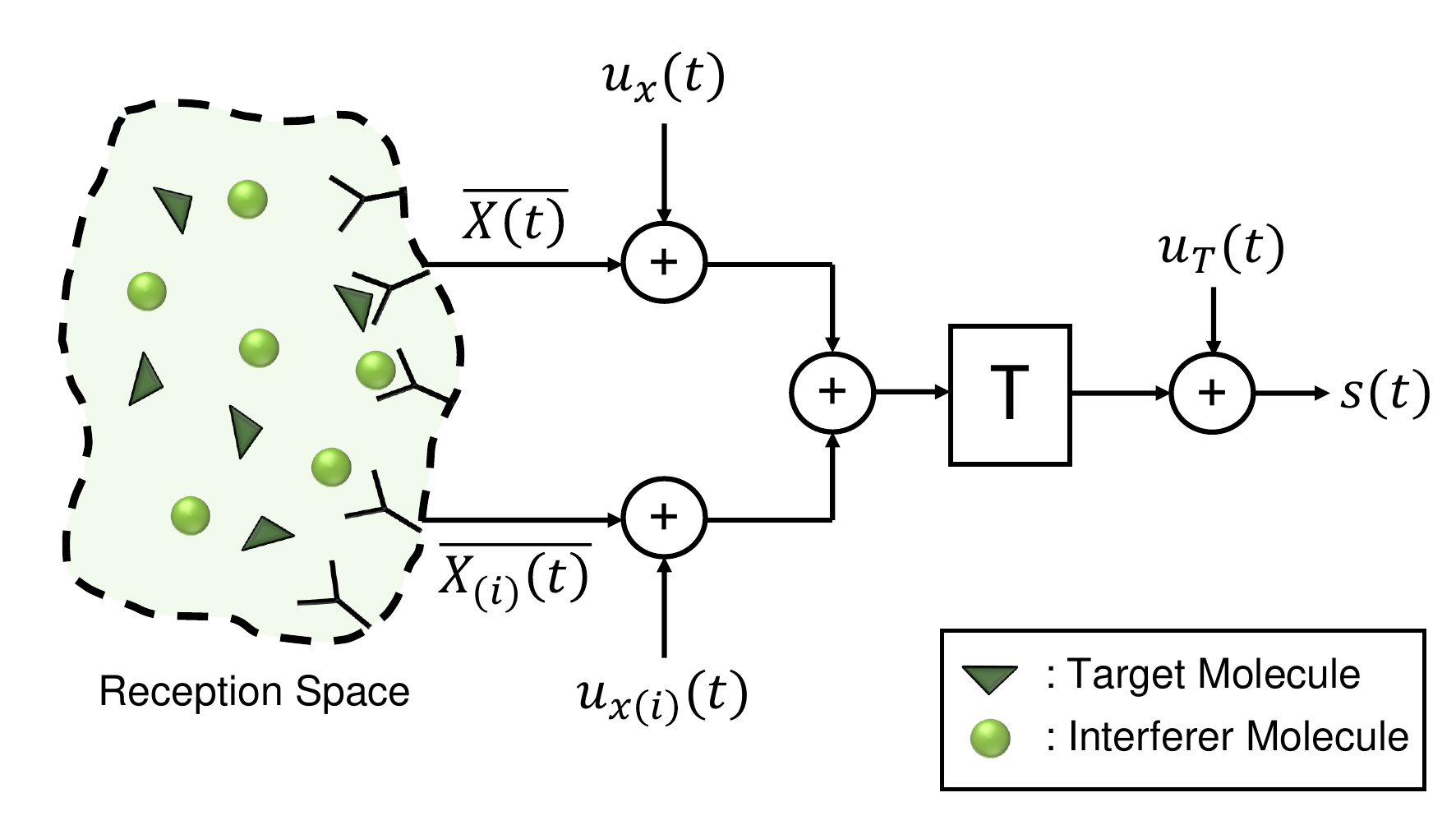}
\caption{Generalized signal flow diagram of a MC-Rx \cite{kuscu2016physical}.}

\label{signalflow}
\end{figure}

\subsubsection{Signal Flow Representation}

To incorporate the effects of competitive binding and molecular interference, the signal flow model of the receiver is depicted in Fig.~\ref{signalflow}. In this model, target molecules (triangles) and interferer molecules (circles) bind to receptor sites in the recognition layer, altering the receptor occupancy and subsequently affecting the transduced signal. The stochastic nature of these interactions introduces fluctuations in the molecular signal, which are further amplified by the transduction process. The electrical output signal can be given by \cite{kuscu2016physical}:
\begin{equation}\label{eq:1}
    s(t) = T\overline{X}(t) + T u_x(t) + u_T (t) + \sum_{i}^{m} T_i \left( \overline{X}_{(i)}(t) + u_{x(i)}(t) \right),
\end{equation}
where \( T \) is the transducing vector that maps receptor occupancy to electrical signal variations, and \( \overline{X}(t) \) represents the ensemble average of the state matrix. The terms \( u_x(t) \) and \( u_T(t) \) denote reception noise and transducing noise, respectively. The contribution of interferer molecules is incorporated through \( \overline{X}_{(i)}(t) \) and its associated reception noise \( u_{x(i)}(t) \), with \( T_i \) representing the transducing response for each interfering species.

The assumption that target and interferer molecules bind independently to receptor sites is a simplification; in reality, competitive binding governs receptor-ligand interactions. The presence of interfering molecules modulates the probability of target binding events, leading to fluctuations in receptor occupancy and, consequently, variations in the transduced signal. This effect becomes particularly significant when receptor availability is limited compared to the combined concentration of target and interferer molecules. In this work, we explicitly incorporate competitive binding dynamics into the Flexure-FET MC receiver model to systematically quantify the impact of molecular interference on signal detection and system performance.

\section{Modeling Flexure-FET-Based MC Receiver for Interference Mitigation via Competitive Binding}

Biological interference in MC systems arises when multiple ligand species compete for the same receptor sites, influencing receptor occupancy and altering the transduced signal. In environments with molecular crowding, this competition introduces fluctuations that affect detection accuracy and system reliability. To systematically characterize these effects, a competitive binding model is introduced, providing a structured framework for analyzing how receptor availability is modulated by the presence of both target and interfering ligands. Unlike conventional approaches that treat interference as external noise, competitive binding inherently shapes the molecular recognition process by dynamically altering receptor occupancy probabilities. In this section, we outline the competitive binding framework and explain its integration into the Flexure-FET MC receiver. We describe how this approach captures the effects of molecular interference and enhances the understanding of receptor dynamics in environments with multiple species. By modeling the competition for receptor sites, the framework improves the receiver's ability to detect molecular signals in high-interference environments.

\subsection{Biorecognition}

In modeling the biorecognition unit, we consider a reaction-limited regime where ligand-receptor interactions dominate over diffusion effects. Under this assumption, the transient fluctuations in ligand concentration can be neglected, allowing the system to be analyzed under steady-state conditions. The receptor layer is exposed to a constant ligand concentration throughout each symbol duration, ensuring that ligand availability remains unchanged for \( t \in [t_i, t_i + 1/B] \), where \( t_i \) represents the transition time between successive messages, and \( 1/B \) is the symbol duration, with \( B \) as the symbol transmission rate. This guarantees a stable receptor-ligand interaction, where the evolution of receptor occupancy is dictated by binding kinetics rather than fluctuations in ligand concentration \cite{kuscu2016physical}.

To accurately model molecular recognition, affinity-based sensing is employed, where receptor-ligand interactions are governed by binding affinities and dissociation constants \cite{rogers2000principles}. Rather than assuming idealized selective binding, we explicitly incorporate competitive binding dynamics, where multiple ligand species interact with the same receptor population based on their relative affinities. Following the framework introduced in \cite{blay2020solving}, receptor occupancy is determined by ligand dissociation constants ($K_j$), with lower values indicating stronger binding affinities. Instead of irreversible absorption, ligand binding is modeled as a temporary and reversible process, where ligands dynamically associate and dissociate from receptor sites. By integrating competitive interactions and affinity-based ranking into the molecular communication model, we achieve a more realistic characterization of molecular interference in the reception process.

\subsubsection{Competitive Binding Framework}

Competitive binding is introduced as a fundamental aspect of receptor-ligand interactions to account for molecular interference in the recognition process. Unlike traditional models that assume isolated binding events, this framework considers the simultaneous interaction of multiple ligand species with the same receptor, leading to competition for binding sites. The approach follows the formalism introduced in \cite{blay2020solving}, which efficiently solves binding equilibria while eliminating the need for initial value estimation. By incorporating ligand competition, this model provides a more realistic characterization of receptor occupancy in molecular environments where multiple species are present.

In this framework, the receptor, denoted as \( P \), binds reversibly with multiple ligand species \( L_1, L_2, \dots, L_n \), forming complexes \( PL_1, PL_2, \dots, PL_n \). The reaction follows a reversible binding scheme where ligands dynamically associate and dissociate with receptor sites \cite{blay2020solving}:

\begin{align}\label{eq:2}
P + L_1 &\rightleftharpoons PL_1 \notag \\
P + L_2 &\rightleftharpoons PL_2 \notag \\
&\vdots \notag \\
P + L_n &\rightleftharpoons PL_n
\end{align}

For a fixed set of initial ligand and receptor concentrations, \( [L_1]_0, [L_2]_0, \dots, [L_n]_0, [P]_0 \), the system reaches a unique equilibrium state as dictated by the deficiency zero theorem \cite{blay2020solving}. 

\subsubsection{Equilibrium Conditions}

At equilibrium, receptor occupancy is governed by the dissociation constant \( K_j \) of each ligand, which quantifies its affinity to the receptor. The fundamental equilibrium condition for each ligand species is given by \cite{blay2020solving}:

\begin{equation}\label{eq:3}
    K_j = \frac{[P]([L_j]_0 - [PL_j])}{[PL_j]}, \quad j = 1,2, \dots, n
\end{equation}

where \( K_j \) represents the dissociation constant for the interaction between \( P \) and \( L_j \). The total receptor balance follows:

\begin{equation}\label{eq:4}
    [P]_0 = [P] + \sum_{j=1}^{n} [PL_j]
\end{equation}

which ensures that all receptors are accounted for in either free or bound states. The equilibrium concentration of each receptor-ligand complex is expressed as:

\begin{equation}\label{eq:5}
    [PL_j] = [L_j]_0 \frac{[P]}{[P] + K_j}, \quad j = 1,2, \dots, n
\end{equation}

where \( [P] \) is the free receptor concentration at equilibrium. Substituting this into the total balance equation leads to the nonlinear equilibrium equation:

\begin{equation}\label{eq:6}
    [P] = [P]_0 - \sum_{j=1}^{n} [L_j]_0 \frac{[P]}{[P] + K_j}
\end{equation}

which can also be rewritten in implicit form:

\begin{equation}\label{eq:7}
    [P]_0 - x - \sum_{j=1}^{n} [L_j]_0 \frac{x}{x + K_j} = 0, \quad \text{where } x = [P]
\end{equation}

To improve numerical stability, an equivalent formulation introduces a quadratic solution for each receptor-ligand complex concentration \cite{blay2020solving}:

\begin{equation}\label{eq:8}
\begin{aligned}
 \relax     [PL_j] = \frac{(P_{\text{avail}} + [L_j]_0 + K_j) - \sqrt{(P_{\text{avail}} + [L_j]_0 + K_j)^2}}{2} \\
    \quad + \frac{- 4 P_{\text{avail}} [L_j]_0}{2}
\end{aligned}
\end{equation}

where \( P_{\text{avail}} \) represents the free receptor available at each iteration. The updated estimate for the available receptor concentration is then given by:

\begin{equation}\label{eq:9}
    P_{\text{avail}} = P_{\text{avail}} + PL_{\text{old}}(j) - PL_{\text{new}}(j)
\end{equation}

\subsubsection{Numerical Solution and Algorithm Implementation}

Due to the nonlinear nature of the equilibrium equations, an iterative algorithm is employed to compute equilibrium concentrations efficiently. The algorithm initializes the available receptor concentration and iteratively updates receptor-ligand complex concentrations until convergence is achieved within a predefined tolerance. The structured algorithm, based on the approach in \cite{blay2020solving}, is provided below.

\begin{algorithm}
\caption{Iterative Equilibrium Solver for Competitive Binding \cite{blay2020solving}}
\begin{algorithmic}[1]
\Require Initial concentrations of receptors ($P_0$), ligands ($L_0$), dissociation constants ($K$)
\Ensure Equilibrium concentrations: Free receptor concentration ($P_{\text{avail}}$), bound complexes ($PL_j$), free ligands ($L_j$)

\State \textbf{Initialize:} Set $P_{\text{avail}}$ and $PL_{\text{old}}$
\State \textbf{Define} convergence tolerance

\While {convergence criterion not met}
    \For {each ligand $j$}
        \State Compute available receptor concentration: $P_{\text{avail}} + PL_{\text{old}}(j)$
        \State Compute new complex: $PL_{\text{new}}(j)$ using (\ref{eq:8})
        \State Update available receptor concentration:
        \[
        P_{\text{avail}} = P_{\text{avail}} + PL_{\text{old}}(j) - PL_{\text{new}}(j)
        \]
    \EndFor
\EndWhile

\State \textbf{Eq. solution given by:}
\State $[P] = P_{\text{avail}}$
\State $[PL_j] = PL_{\text{new}}(j)$
\State $[L_j] = L_0(j) - PL_{\text{new}}(j)$

\State \textbf{Return:} $P_{\text{avail}}, PL_{\text{new}}, L_j$

\end{algorithmic}
\end{algorithm}

\subsection{Binding Noise}

Surface receptors engage in temporary and reversible ligand binding, randomly interacting with ligands present in their vicinity. Each receptor transitions between the bound and unbound states based on the kinetics of ligand-receptor interactions. The transition between these states is governed by the local ligand concentration and the receptor’s binding and unbinding rates \cite{kuscu2019sensing}. 

When multiple ligand species coexist in the environment and compete for binding to the same receptor, each ligand type exhibits a different dissociation constant, influencing the receptor occupancy. Under these conditions, the probability of a receptor being in the bound state at equilibrium is given by \cite{kuscu2019sensing}:

\begin{equation} \label{eq:10}
    p_B = \frac{\sum_{j=1}^{n} \frac{[L_j]_0}{K_j}}{1 + \sum_{j=1}^{n} \frac{[L_j]_0}{K_j}},
\end{equation}

where \( n \) denotes the number of ligand species present, \( [L_j]_0 \) represents the total concentration of ligand type \( j \) in the reception space at the sampling time, and \( K_j = k_j^- / k_j^+ \) is its respective dissociation constant, with \( k_j^+ \) and \( k_j^- \) denoting the binding and unbinding rate constants of ligand type \( j \), respectively.

If receptors are independently exposed to multiple ligand species and do not interact with one another, the number of bound receptors follows a binomial distribution, with variance expressed as \cite{kuscu2016physical}:

\begin{equation} \label{eq:11}
    \text{Var}(N_B) = p_B (1 - p_B) N_R.
\end{equation}

Here, \( N_R \) represents the total number of receptors, given by \( N_R = A [P]_0 \), where \( [P]_0 \) denotes the concentration of surface receptors. The gate area is defined as \( A = W \times L \), where \( W \) and \( L \) correspond to the beam width and beam length, respectively. These expressions characterize the binding noise in MC systems, arising due to the stochastic nature of ligand-receptor interactions. Furthermore, in competitive binding scenarios, where multiple ligand types interact with the same receptor population, receptor occupancy fluctuations are further influenced by differential binding affinities, contributing to variations in the detected signal.

The presence of multiple ligand species introduces additional complexity to receptor occupancy dynamics, as each ligand type binds with a different affinity, contributing to fluctuations in receptor state transitions. In such cases, the total binding and unbinding rates are no longer determined by a single ligand species but instead by the combined effects of all competing ligands.

The total binding rate, accounting for the contributions of all ligand species, is given by:

\begin{equation}\label{eq:12}
    k_{\text{on,total}} = \sum_{j} k_j^+ [L_j]_0.
\end{equation}

Similarly, the total unbinding rate, weighted by the probability that a receptor is occupied by ligand \( j \), is expressed as:

\begin{equation}\label{eq:13}
    k_{\text{off,total}} = \sum_{j} p_{B,j} k_j^-,
\end{equation}

where the probability of receptor occupancy by ligand \( j \) is:

\begin{equation}\label{eq:14}
    p_{B,j} = \frac{\frac{[L_j]_0}{K_j}}{1 + \sum_{n} \frac{[L_j]_0}{K_n}}.
\end{equation}

Since receptor state transitions follow a first-order kinetic process, the characteristic binding noise relaxation time, governing how quickly receptor occupancy fluctuations return to equilibrium, is given by:

\begin{equation}\label{eq:15}
    \tau_{\text{multi}} = \frac{1}{\sum_{j} k_j^+ [L_j]_0 + \sum_{j} p_{B,j} k_j^-}.
\end{equation}

Receptor occupancy fluctuations due to stochastic ligand binding introduce time-correlated noise into the molecular communication system. The autocorrelation function of these fluctuations follows an exponential decay \cite{kuscu2016physical}:

\begin{equation}\label{eq:16}
    R(\tau) = \text{Var}(N_B) e^{-\frac{\tau}{\tau_B}}.
\end{equation}

The power spectral density (PSD), derived from the Fourier Transform of \( R(\tau) \), is given by \cite{kuscu2016physical}:

\begin{equation}\label{eq:17}
    S_{N_B}(f) = \frac{\text{Var}(N_B)}{2 \tau_B} \frac{1}{1 + (2 \pi f \tau_B)^2}.
\end{equation}

This expression characterizes the frequency-dependent behavior of receptor binding noise. At lower frequencies, fluctuations in receptor occupancy contribute significantly to noise, affecting signal detection performance, whereas at higher frequencies, the impact of binding noise diminishes.

By incorporating the effects of competitive ligand binding into the receptor noise model, this formulation provides a quantitative framework for analyzing how ligand-ligand interactions influence reliability. The extended model accounts for the interplay between receptor-ligand interactions, competitive binding, and stochastic fluctuations, offering insights into noise mitigation strategies for MC receivers.

\subsection{Transducer}

\begin{figure}[t]
  \centering
  \includegraphics[width=0.6\linewidth]{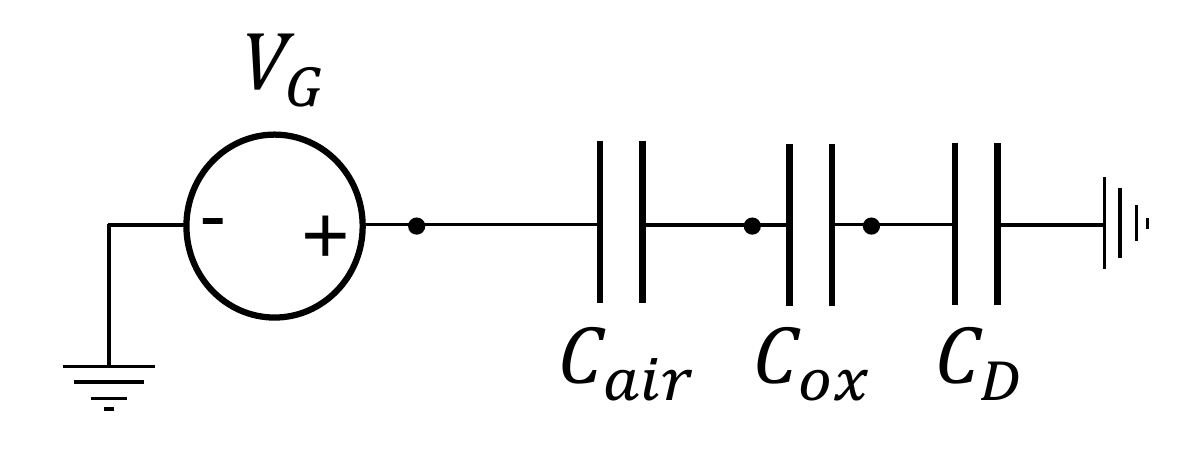}
  \caption{The equivalent circuit of Flexure-FET MC-Rx.}
  \label{circuit}
\end{figure}

As depicted in Fig. \ref{circuit}, the Flexure-FET operates through three fundamental mechanisms: (\textit{i}) alteration of mechanical stiffness due to molecular binding (\(\Delta k\)), (\textit{ii}) application of bias near the pull-in threshold to optimize gate displacement (\(\Delta y\)), and (\textit{iii}) subthreshold conduction, which enables an exponential change in drain current \cite{jain2012flexure}. The interaction between these mechanical and electrical properties determines the overall response of the device, where ligand binding at the gate surface induces changes in its displacement and electrical characteristics.

The equivalent circuit of the Flexure-FET is shown in Fig. \ref{circuit}, where \( C_{\text{air}} \) is the capacitance of the air gap, \( C_{\text{ox}} \) is the capacitance of the oxide layer, and \( C_D \) is the depletion capacitance. Among these, \( C_{\text{air}} \) varies dynamically as ligand binding alters the gate position, affecting the electrostatic forces and, consequently, the electrical response of the receiver.

The mean number of bound ligands is represented by \( \mu_{N_s} \), accounting for both target and interferer molecules. The contribution from the target molecules is given by:

\begin{equation}\label{eq:18}
    \mu_{N_s^T,m} = PL_1
\end{equation}

where \( \mu_{N_s^T,m} \) represents the mean bound receptor density due to target molecules for symbol \( m \). The concentration \( [L_j]_0 \) refers to the total ligand concentration in the reception space at the sampling time, including both target and interferer molecules. This measurement is taken at the end of the signaling interval, corresponding to the reception of symbol \( m \), where \( m \in \{0, 1, ..., M-1\} \) for the \( k \)th time slot.

Similarly, the contribution from interferer molecules is:

\begin{equation}\label{eq:19}
    \mu_{N_s^I,m} = \sum_{n=2}^{n} PL_n.
\end{equation}

where \( \mu_{N_s^I,m} \) represents the mean bound receptor density due to interfering ligands. Since the binding process is reversible, ligands continuously associate and dissociate from receptor sites, contributing to fluctuations in receptor occupancy. Here, \( PL_1 \) corresponds to the equilibrium concentration of bound target molecules, while \( PL_{2,n} \) represents the equilibrium-bound interferer molecules for each species \( n \).

Thus, the total bound receptor density at time \( kT_s \) is given by combining (\ref{eq:18}) and (\ref{eq:19}):

\begin{equation}\label{eq:20}
    \mu_{N_{sum},m} = \mu_{N_s^T,m} + \mu_{N_s^I,m}.
\end{equation}

The bound receptor densities for target and interferer molecules can be expressed in a compact vector form as:

\begin{equation}\label{eq:21}
    \boldsymbol{\mu}_{N_s,m} =
    \begin{bmatrix}
        \mu_{N_s^T,m} \\
        \mu_{N_s^I,m} \\
        \mu_{N_{\text{sum}},m}
    \end{bmatrix}.
\end{equation}

The corresponding mean values of the stiffness variation \( \boldsymbol{\mu}_{\Delta {k_m}} \), gate displacement \( \boldsymbol{\mu}_{\Delta y_m} \), surface potential shift \( \boldsymbol{\mu}_{\Delta \psi_m} \), and the sensitivity \( \boldsymbol{\mu}_{S_m} \), defined as the ratio of drain currents before and after ligand binding, can be obtained. The stiffness change due to molecular interactions at the gate is modeled based on an effective thickness variation of the flexible structure. Assuming a uniform distribution of bound ligands across the sensor surface, the mean stiffness change is approximated as \cite{jain2012flexure}:

\begin{equation}\label{eq:22}
   \frac{\boldsymbol{\mu}_{\Delta k_m}}{k} \approx \frac{3\boldsymbol{\mu}_{N_s,m}MV_m}{H},
\end{equation}

where \( k \) is the mechanical stiffness, \( H \) is the gate thickness, and \( MV_m \) is the molecular volume of the bound ligands, which can be determined from their molecular weight \( MW_m \).

The displacement of the gate, denoted by \( y \), is approximately given as \( y \approx \frac{2}{3}y_0 \) when operating near the pull-in threshold, where \( y_0 \) is the initial separation between the gate and the substrate. The static equilibrium of the Flexure-FET is established by balancing spring and electrostatic forces. The relationship between the gate displacement and the stiffness variation induced by ligand binding is expressed as \cite{jain2012flexure}:

\begin{equation}\label{eq:23}
 (3y - y_0) \Delta y^2 + y(3y - 2y_0) \Delta y  \approx \frac{\epsilon_0A(V_G - \psi_s)^2}{2} \frac{\Delta k}{k^2}
\end{equation}

where \( \epsilon_0 \) is the permittivity of free space, \( \Delta y \) represents the change in gate position, and \( \psi_s \) is the surface potential. Using (\ref{eq:21}) and (\ref{eq:22}), the mean gate displacement is given by:

\begin{equation}\label{eq:24}
  \boldsymbol{\mu}_{\Delta y_m}  \approx \sqrt{\frac{\epsilon_0 A (V_G - \psi_s)^2}{2(3y - y_0)} \frac{\boldsymbol{\mu}_{N_s,m}MV_m}{Hk}}.
\end{equation}

The mean change in surface potential is then expressed as:

\begin{equation}\label{eq:25}
  \boldsymbol{\mu}_{\Delta \psi_m} \approx \frac{-k \boldsymbol{\mu}_{\Delta y_m} + \boldsymbol{\mu}_{\Delta k_m}(y_0 - y)}{q \epsilon_s N_A A},
\end{equation}

where \( q \) is the elementary charge, \( \epsilon_s \) is the dielectric constant of the substrate, and \( N_A \) is the doping concentration in the substrate. The mean sensitivity of the device, which quantifies the impact of ligand binding on the drain current, is given by:

\begin{equation}\label{eq:26}
    S = \frac{I_{DS1}}{I_{DS2}}  \approx \exp\biggl(\frac{k \boldsymbol{\mu}_{\Delta y_m} - \boldsymbol{\mu}_{\Delta k_m}(y_0 - y)}{k_B T \epsilon_s N_A A}\biggl),
\end{equation}

where \( I_{DS2} = \boldsymbol{\mu}_{I_m} \) is the mean output current, \( k_B \) is the Boltzmann constant, and \( T \) is the absolute temperature. The mean output current is represented as \( \boldsymbol{\mu}_{I_m} = [\mu_{I_T,m}, \mu_{I_I,m}, \mu_{I_{\text{sum}},m}]^\top \), where \( \mu_{I_T,m} \) and \( \mu_{I_I,m} \) denote the contributions from target and interferer molecules, respectively, and \( \mu_{I_{\text{sum}},m} \) is their sum. This formulation enables independent processing of each contribution while preserving their combined effect.

Flicker noise, also known as \( 1/f \) noise, is the primary noise component influencing transduction in the Flexure-FET. Other noise sources, such as thermomechanical force noise, which results from stochastic gate position fluctuations due to molecular interactions, contribute negligibly in comparison to flicker and binding noise and are therefore considered negligible in this work \cite{aktas2021mechanical}. The behavior of flicker noise follows the correlated carrier number and mobility fluctuation model, with its power spectral density (PSD) given by:

\begin{equation}\label{eq:27}
    S_{I^F_m}(f) = \frac{\lambda k_B T q^2 N_{ot} g_{\text{FET}}^2}{W L C_{\text{ox}}^2 |f|} \left[1 + \alpha_s \mu_p C_{\text{ox}} (V_G - |V_{TH}|) \right]^2,
\end{equation}

where \( \lambda \) denotes the characteristic tunneling distance, \( N_{ot} \) represents the oxide trap density, and \( g_{\text{FET}} = \frac{\partial I_{DS}}{\partial V_G} \) corresponds to the transconductance of the FET. Furthermore, \( \alpha_s \) is the Coulomb scattering coefficient, \( \mu_p \) represents the hole carrier mobility, and \( V_{TH} \) is the FET’s threshold voltage \cite{rajan2010temperature}.

The overall noise in the system is influenced by both the binomial distribution of bound receptors and the flicker noise, which, for a sufficiently high number of surface receptors (\(N_R > 1000\)), can be approximated as a Gaussian distribution \cite{hooge1969amplitude}. As a result, the total noise in the system is considered to be the sum of two independent Gaussian noise processes, as described in \cite{kuscu2016modeling, aktas2021mechanical}. The power spectral density (PSD) of the total noise that affects the output current is given by:

\begin{equation} \label{eq:28}
    S_{I_m}(f) = S_{I_m^{B}}(f) + S_{I_m^{F}}(f),
\end{equation}

where \( S_{I_m^{B}}(f) = S_{N_{B,m}}(f) \times \Delta \psi^2 \times g_{\text{FET}}^2 \), with \( \Delta \psi \) representing the change in surface potential due to a single bound ligand. Additionally, \( S_{N_{B,m}}(f) \) can be computed using (\ref{eq:17}) for the corresponding symbol. The variance of the output current, \( \sigma_{I_m}^2 \), is determined by integrating the total noise power spectral density (\ref{eq:17}) over all frequencies \cite{kuscu2016physical}:

\begin{equation} \label{eq:29}
   \sigma_{I_m}^2 = \int_{-\infty}^{\infty} S_{I_m}(f) \, df.
\end{equation}

\section{Performance Analysis}

In this section, we present numerical results based on the model introduced in the previous section to evaluate the performance of the Flexure-FET MC receiver in interference-rich environments. Specifically, we compare the receiver's performance in scenarios that consider the effects of interference through competitive binding, as introduced earlier, with those where interference is not accounted for. We utilized the previously improved design of the Flexure-FET MC receiver from \cite{aktas2022weight}, with dimensions chosen to be $W = 1~\mu m$, $L = 4~\mu m$, $H = 40~\text{nm}$, $y_0 = 100~\text{nm}$, and $E = 200~\text{GPa}$. Additionally, we used a molecule set with molecular weights ranging from $89~\text{g/mol}$ to $763~\text{g/mol}$. We assumed that interferer molecules have the same molecular weight as the target molecules. The default parameter values for the simulation scenarios are provided in Table \ref{tab:para}. These values were selected based on commonly used parameters in the MC literature \cite{kuscu2016modeling, kuscu2016physical}.

\begin{figure}[!t]
  \centering
  \includegraphics[width=0.9\linewidth]{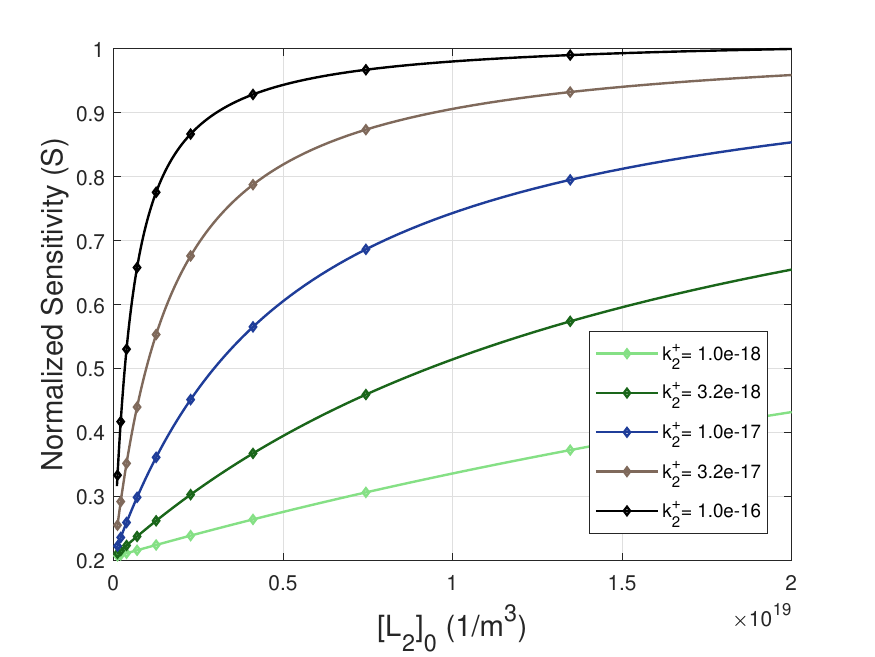}
  \caption{Normalized Sensitivity (S) as a function of interferer molecule concentration \( [L_2]_0 \) for different \( k_2^+ \) values. }
  \label{sens}
\end{figure}

\begin{table}
  \caption{Simulation Parameters}
  \label{tab:para}
  \begin{tabular}{ll}
    \toprule
    \midrule

Concentration of surface receptors  $([P]_0)$ & $5 \times 10^{18}\:m^{-2} $\\
Binding rate  $(k_1)$  & $3 \times 10^{-18} \:  m^{3}/s $\\
Unbinding rate  $(k_{-1})$  & $20\:  s^{-1}  $\\
Young modulus of beam  $(E)$  & $4    \:  GPa$\\
Beam length  $(L)$          & $8     \:  \mu m $\\
Beam width  $(W)$          & $1    \:  \mu m $\\
Beam thickness $(H)$          & $260    \:  nm $\\
Air gap $(y_0)$          & $100    \:  nm $\\
Substrate doping $(N_A)$          & $10^{16}    \:  cm^{-3} $\\
Dielectric thickness $(y_d)$          & $10   \:  nm $\\
Oxide trap density  $(N_{ot})$  & $2.3 \times 10^{24} \:  eV^{-1}cm^{-3} $\\
  \bottomrule
\end{tabular}
\end{table}
\subsection{Sensitivity \& Noise}

First, we present a comparative analysis of the receiver response, focusing on the receiver's sensitivity as a function of interferer molecule concentration \( [L_2]_0 \), considering different system parameters. In Fig. \ref{sens}, the Normalized Sensitivity (S) is plotted against interferer molecule concentration \( [L_2]_0 \) for varying \( k_2^+ \) values. The sensitivity values are normalized to the maximum sensitivity value observed. As shown in the figure, sensitivity increases with the concentration of interferer molecules but eventually reaches a saturation point. Beyond this threshold, further increases in concentration do not significantly enhance sensitivity.

This behavior highlights the importance of \( k_2^+ \) values, as they directly influence the rate at which the system responds to increasing interferer concentration. Higher \( k_2^+ \) values result in a faster increase in sensitivity at lower concentrations, allowing the receiver to more effectively detect interferers. However, for lower \( k_2^+ \) values, the sensitivity increases more gradually, indicating slower binding kinetics and a less efficient response to rising concentrations. While higher concentrations improve sensitivity at lower values of \( [L_2]_0 \), there is an optimal concentration beyond which the system's sensitivity no longer improves, emphasizing the importance of understanding the interplay between interferer concentration and binding rates for effective system performance.

\begin{figure}[!t]
  \centering
  \includegraphics[width=0.8\linewidth]{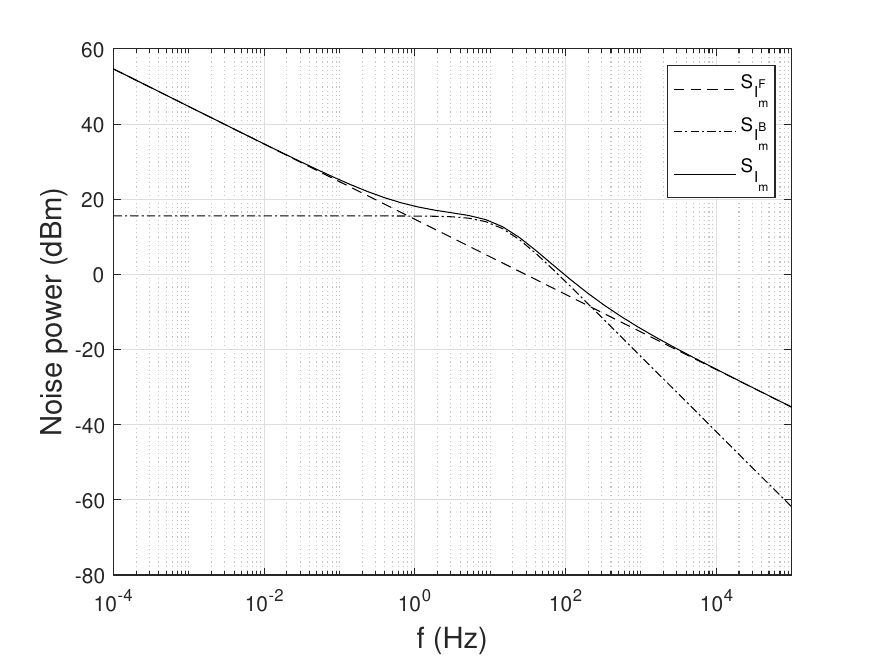}
\caption{Individual contributions of noise sources to the MC receiver: \( S_{I_m^{B}} \) representing binding noise, \( S_{I_m^{F}} \) representing flicker noise, and \( S_{I_m} \) representing total noise.}
  \label{fig:noise}
\end{figure}

In Fig. \ref{fig:noise}, the contributions of different noise sources to the MC receiver's output current are shown. The plot includes \( S_{I_m^{B}} \) (binding noise), \( S_{I_m^{F}} \) (flicker noise), and \( S_{I_m} \) (total noise). At lower frequencies, binding noise dominates, primarily due to the randomness in ligand-receptor binding events. As the frequency increases, flicker noise becomes more significant, reflecting fluctuations in voltage that affect the receiver. The total noise, \( S_{I_m} \), is the combined effect of both types of noise. This plot demonstrates how each noise source contributes at different frequencies and affects the receiver’s overall performance. Understanding these noise behaviors is crucial for improving system design, as minimizing the combined noise impact, especially at critical frequencies, is key to enhancing the receiver's ability to accurately detect molecular signals.

\begin{figure*}[!t]
\centering
\subfloat[]{\includegraphics[width=0.60\columnwidth]{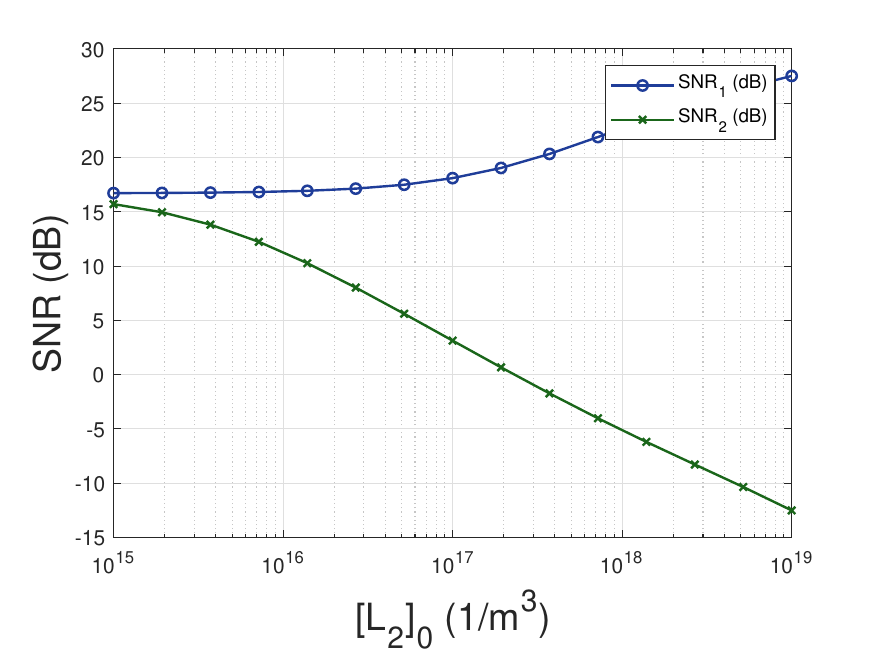}%
\label{fig_first_case}}
\hfil
\subfloat[]{\includegraphics[width=0.60\columnwidth]{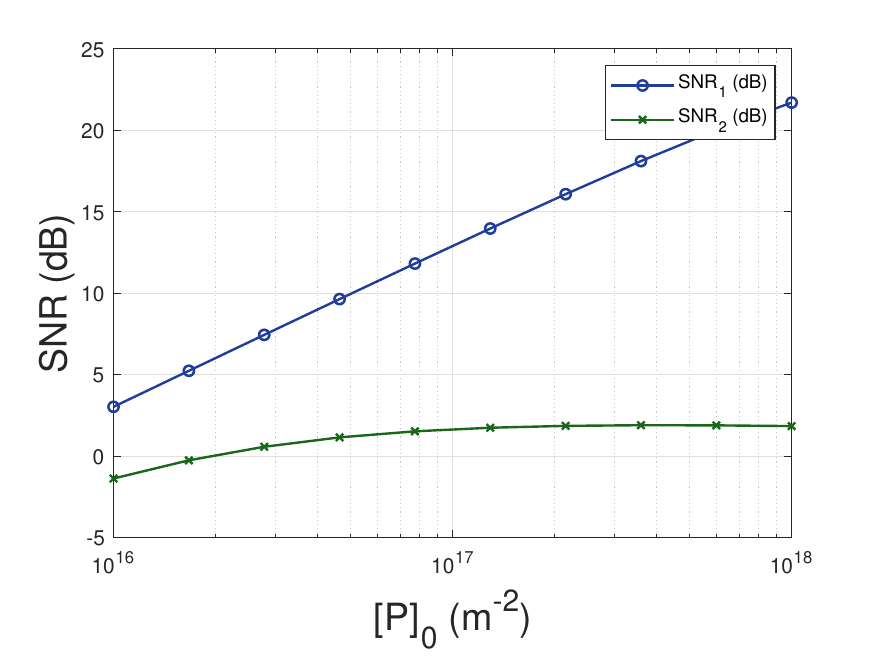}%
\label{fig_second_case}}
\hfil
\subfloat[]{\includegraphics[width=0.60\columnwidth]{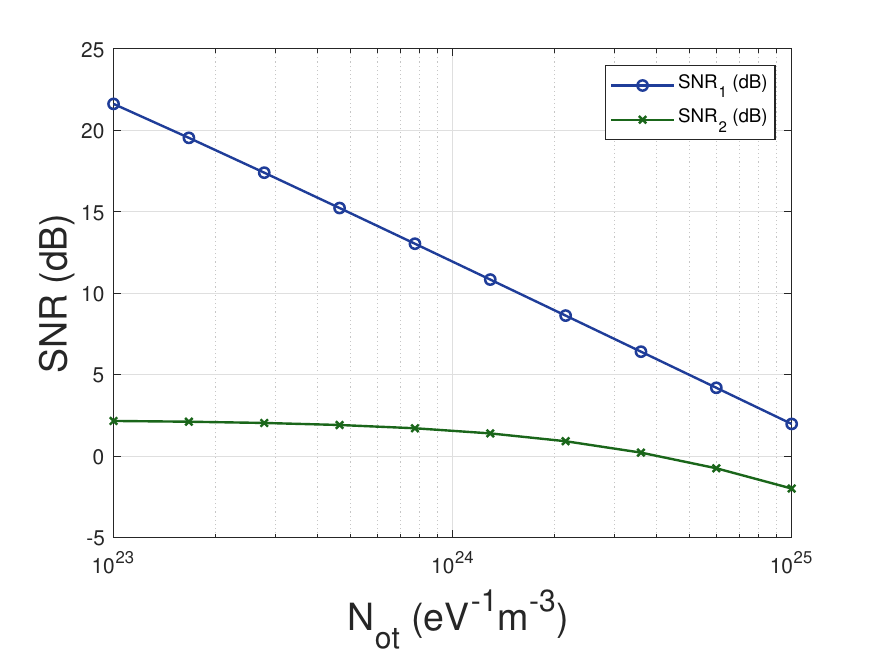}%
\label{fig_third_case}}
\hfil
\subfloat[]{\includegraphics[width=0.60\columnwidth]{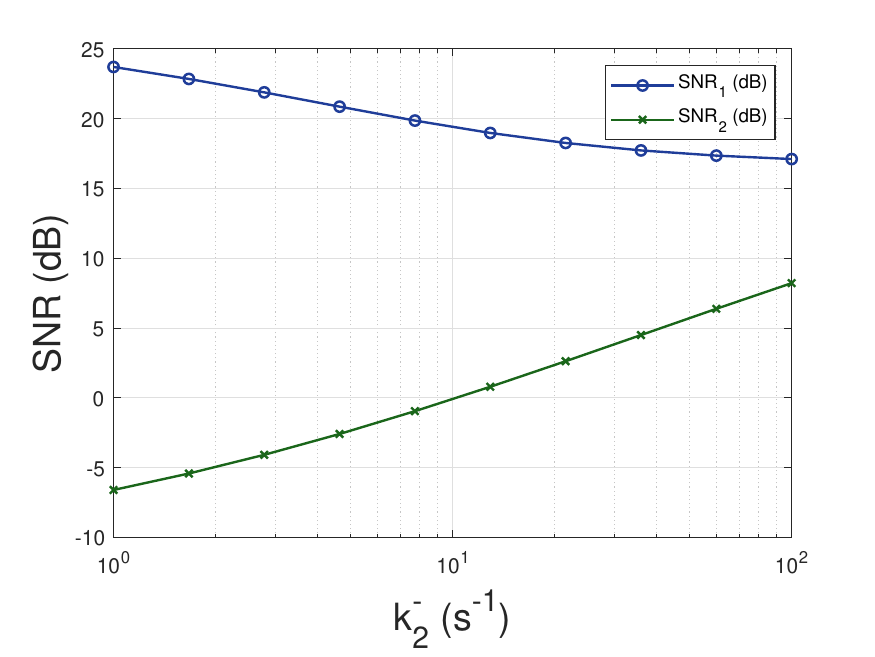}%
\label{fig_fourth_case}}
\hfil
\subfloat[]{\includegraphics[width=0.60\columnwidth]{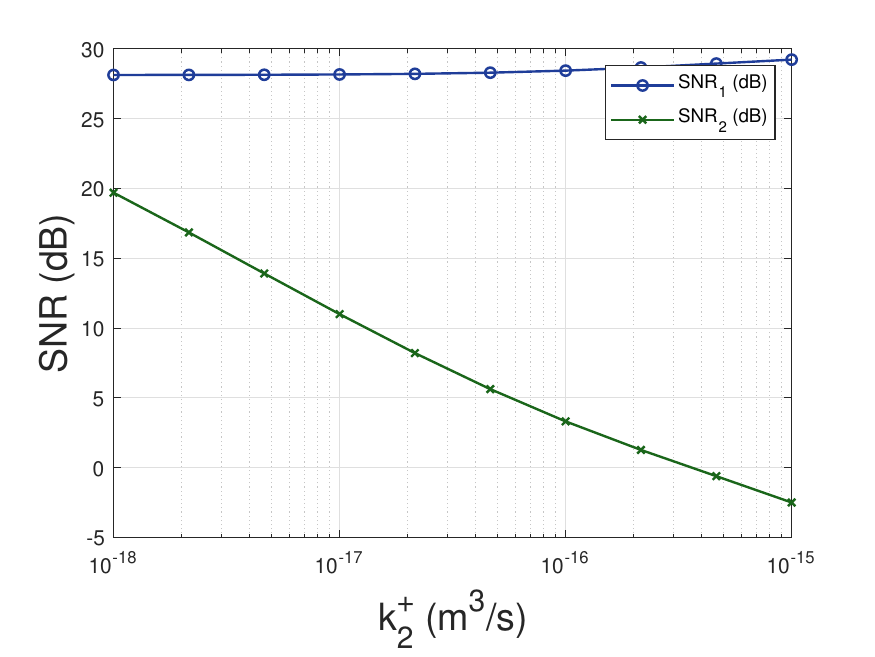}%
\label{fig_fifth_case}}
\hfil
\caption{SNR as a function of (a) interferer molecule concentration \( [L_2]_0 \), (b) surface receptor concentration \( [P_0] \), (c) oxide trap density \( N_{\text{ot}} \), (d) interferer binding rate constant \( k_2^{+} \), and (e) the interferer unbinding rate constant \( k_2^{-} \), for each type of SNR formulation: \( \text{SNR}_1 \) (blue) and \( \text{SNR}_2 \) (green).} 
\label{snr1}
\end{figure*}

\begin{figure}[!t]
  \centering
  \includegraphics[width=0.8\linewidth]{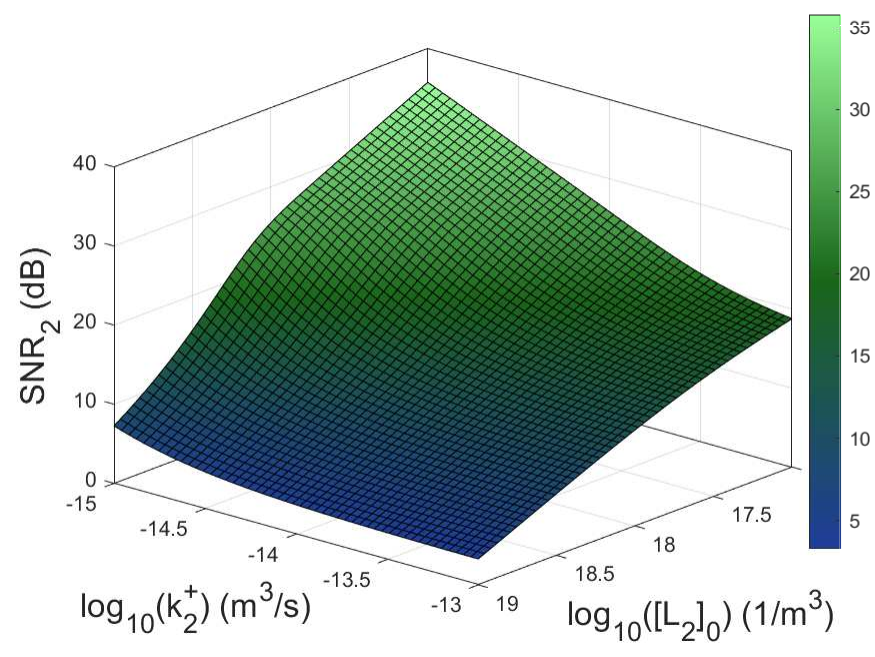}
\caption{3D plot of SNR\(_2\) vs. \( [L_2]_0 \) and \( k_2^+ \) on log scale, showing the effect of interferer concentration and binding rate on SNR.}

  \label{snr_3d}
\end{figure}

\begin{figure*}[!t]

\centering
\subfloat[]{\includegraphics[width=0.6\columnwidth]{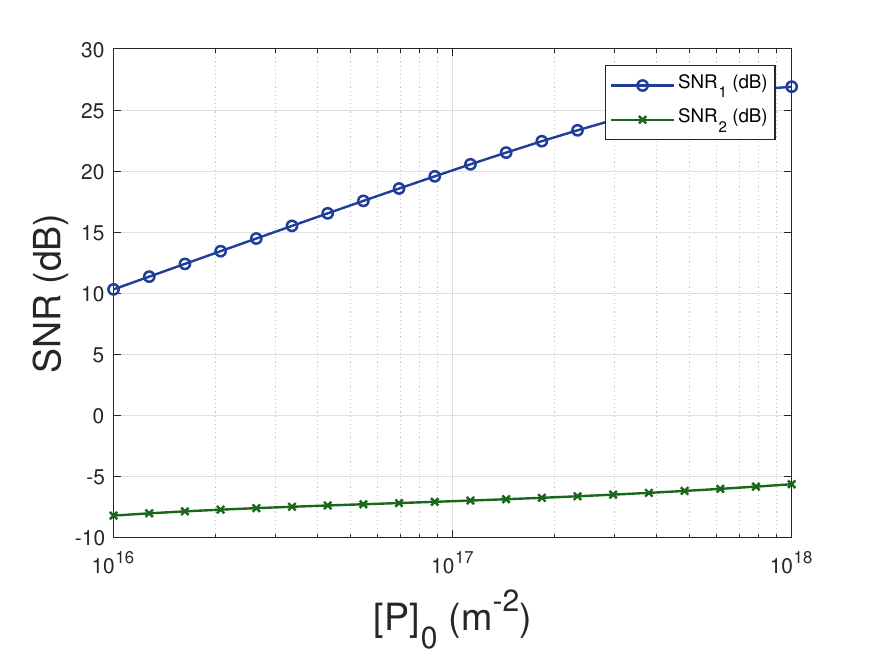}%
\label{fig_first_casee}}
\hfil
\subfloat[]{\includegraphics[width=0.6\columnwidth]{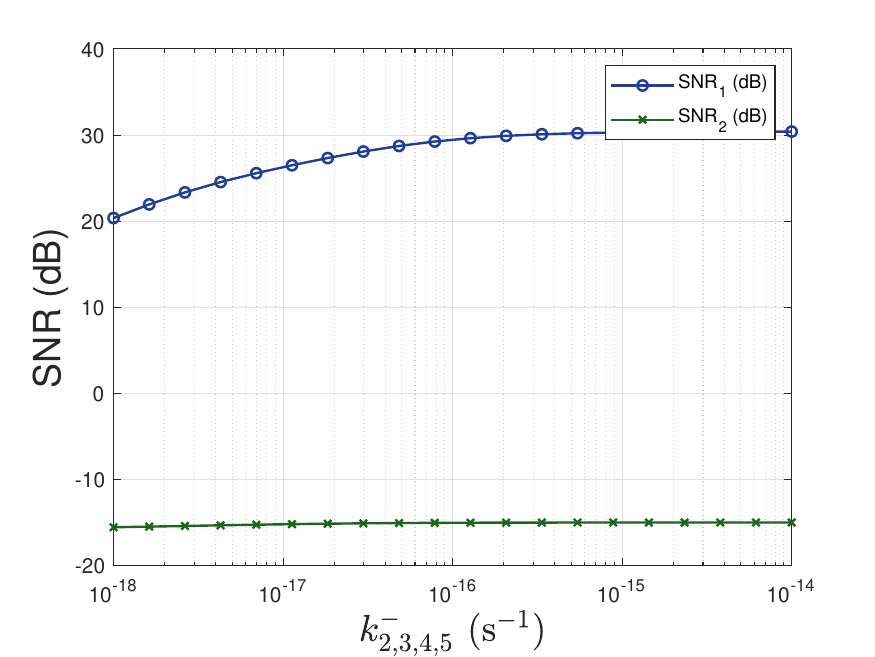}%
\label{fig_second_casee}}
\hfil
\subfloat[]{\includegraphics[width=0.6\columnwidth]{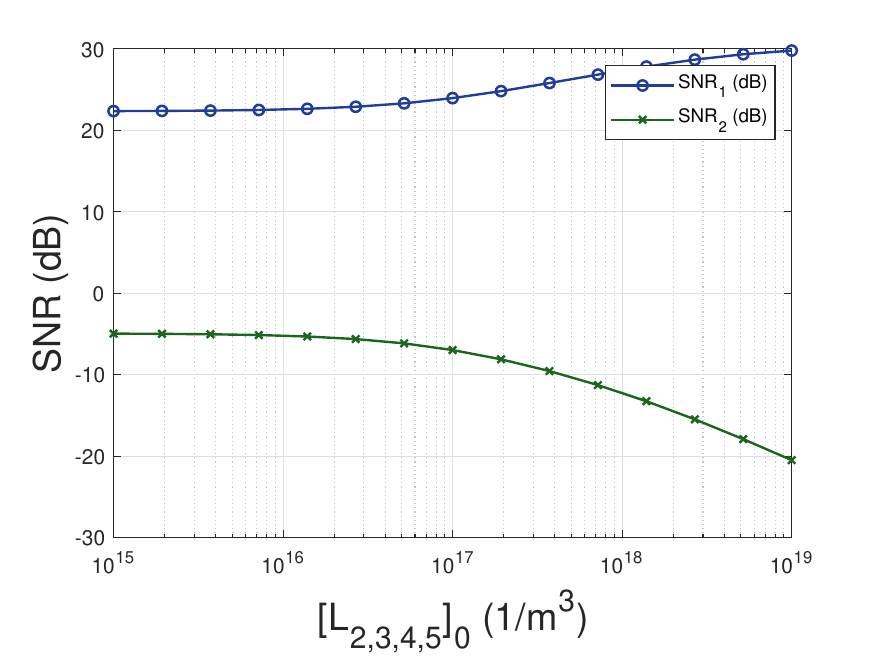}%
\label{fig_third_casee}}
\hfil
\caption{SNR as a function of (a) surface receptor concentration \( [P_0] \), (b) interferer binding rate constant \( k_{2,3,4,5}^+ \), and (c) interferer molecule concentration \([L_{2,3,4,5}]_0 \) for multiple interferer molecules.
}
\label{snr_multi}
\end{figure*}

\subsection{SNR Analysis}

The output signal-to-noise ratio (SNR) is calculated under varying system parameters, including communication system and receiver design factors. To examine the impact of interferer molecules in competitive binding scenarios, two distinct formulations of SNR are defined.

The first formulation, \( \text{SNR}_{1} \), considers the system noise without taking into account interference molecules as noise. It represents the SNR in a scenario where noise arises solely from intrinsic system sources. The formula for \( \text{SNR}_{1} \) is:

\begin{equation}\label{eq:30}
      \text{SNR}_{1,m} = \frac{\mu_{I_{\text{sum}},m}^2}{\sigma_{I_m}^2},
\end{equation}

where \( \mu_{I_{\text{sum}},m} \) is the total contribution from both target and interferer molecules, and \( \sigma_{I_m}^2 \) is the variance of the current output.output.

The second formulation, \( \text{SNR}_{2} \), incorporates the interference molecules as part of the noise, providing a more accurate reflection of the performance of the system under real-world conditions. This is expressed as follows:

\begin{equation}\label{eq:31}
      \text{SNR}_{2,m} = \frac{\mu_{I_T,m}^2}{\sigma_{I_m}^2 + \mu_{I_I,m}^2},
\end{equation}

where \( \mu_{I_T,m} \) and \( \mu_{I_I,m} \) represent the contributions from target and interferer molecules, respectively, and the denominator reflects the combined noise variance from both the intrinsic system noise and the noise due to interferers.

By comparing \( \text{SNR}_{1} \) and \( \text{SNR}_{2} \), the influence of interferer molecules on the performance of the system is analyzed in competitive binding scenarios. This comparison highlights the importance of considering interferer molecules as part of the noise, providing a more realistic understanding of the system’s performance in real-world conditions. Incorporating interferer molecules into the noise model (i.e., \( \text{SNR}_2 \)) allows for a more comprehensive assessment of the system’s sensitivity to noise and its ability to discriminate signals amidst biological interference. These insights are crucial for optimizing the system’s response, and by fine-tuning ligand concentrations or receptor affinities, competitive binding can be leveraged to enhance the biosensor's ability to detect specific molecules in noisy environments, making it possible to optimize the performance for practical applications.

In Fig. \ref{snr1}(a), the comparison between \( \text{SNR}_1 \) and \( \text{SNR}_2 \) as a function of interferer molecule concentration \( [L_2]_0 \) is shown. As expected, \( \text{SNR}_1 \) remains relatively stable at higher concentrations of \( [L_2]_0 \), implying that noise is primarily dominated by intrinsic factors. However, \( \text{SNR}_2 \) decreases as \( [L_2]_0 \) increases, reflecting the contribution of interferer molecules to the noise, which leads to a decrease in SNR. This highlights the significance of considering interferers as part of the noise, offering a more accurate assessment of the receiver's performance in real-world scenarios.

Fig. \ref{snr1}(b) shows the effect of surface receptor concentration \( [P_0] \) on the SNR. While \( \text{SNR}_1 \) improves with increasing \( [P_0] \), \( \text{SNR}_2 \) remains constant, suggesting that while receptor concentration enhances signal strength, it does not significantly affect the noise from interferers. This emphasizes the potential for tuning the receiver's response by adjusting receptor concentrations to enhance sensitivity without amplifying the interference noise.

\subsubsection{Single Interferer Case}
In Fig. \ref{snr1}(c), the relationship between \( \text{SNR}_1 \) and \( \text{SNR}_2 \) as a function of oxide trap density \( N_{\text{ot}} \) is shown. As \( N_{\text{ot}} \) increases, both \( \text{SNR}_1 \) and \( \text{SNR}_2 \) decrease. The blue curve, \( \text{SNR}_1 \), decreases more sharply, indicating that the intrinsic system noise is more significantly impacted by the increasing oxide trap density. On the other hand, \( \text{SNR}_2 \) (the green curve) also decreases with increasing \( N_{\text{ot}} \), but the rate of decline is less severe, reflecting the fact that the noise contribution from interferer molecules remains relatively constant despite the increase in oxide trap density. This behavior emphasizes the impact of oxide trap density on system performance, particularly how it increases intrinsic noise. While the contribution from interferers remains relatively stable, the increased oxide trap density amplifies the overall noise, leading to a decrease in both SNR formulations.

Fig. \ref{snr1}(d) shows how the interferer binding rate constant \( k_2 \) affects the SNR. As \( k_2 \) increases, both \( \text{SNR}_1 \) and \( \text{SNR}_2 \) decrease. The faster binding of interferer molecules leads to more frequent binding events, increasing noise and reducing sensitivity. This demonstrates how the binding kinetics of interferers can be controlled to optimize the receiver’s performance in competitive binding scenarios.

Fig. \ref{snr1}(e) analyzes the effect of the interferer unbinding rate constant \( k_2^{-} \) on the SNR. While \( \text{SNR}_1 \) remains stable, \( \text{SNR}_2 \) decreases as \( k_2^{-} \) increases. Faster unbinding of interferers increases the overall noise, highlighting the need to manage unbinding rates to minimize interference and optimize receiver sensitivity.

In Fig. \ref{snr1}(d), the relationship between \( \text{SNR}_1 \) and \( \text{SNR}_2 \) as a function of the unbinding rate \( k_2^{-} \) is shown. As \( k_2^{-} \) increases, \( \text{SNR}_1 \) decreases slightly due to increased intrinsic noise. However, \( \text{SNR}_2 \) increases because more frequent unbinding events of interferer molecules reduce their noise contribution, improving the signal detection and overall sensitivity.

In Fig. \ref{snr1}(e), the effect of the interferer binding rate constant \( k_2^{+} \) on \( \text{SNR}_1 \) and \( \text{SNR}_2 \) is shown. As \( k_2^{+} \) increases, \( \text{SNR}_1 \) remains stable, since it does not account for the noise from interferers. In contrast, \( \text{SNR}_2 \) decreases as faster binding of interferers increases their noise contribution, which reduces the system’s sensitivity. This highlights the importance of controlling the interferer binding rate to optimize performance.

In Fig. \ref{snr_3d}, the surface plot shows \( \text{SNR}_2 \) as a function of interferer binding rate \( k_2^+ \) and interferer concentration \( [L_2]_0 \). As \( [L_2]_0 \) increases, \( \text{SNR}_2 \) decreases, indicating higher interference. Similarly, \( \text{SNR}_2 \) decreases with higher \( k_2^+ \), suggesting more interference from higher binding rates. The plot demonstrates that the system is more sensitive to changes in ligand concentration than in binding rate due to the broader impact of ligand concentration on system behavior.

\subsubsection{Multi-Interferer Case}
In Fig. \ref{snr_multi}(a), the effect of surface receptor concentration \( [P_0] \) on the SNR is shown for the case with two interferer molecules. As \( [P_0] \) increases, both \( \text{SNR}_1 \) and \( \text{SNR}_2 \) improve, but the decrease from \( \text{SNR}_1 \) to \( \text{SNR}_2 \) is more significant compared to Fig. \ref{snr1}(b). This is expected because, as the concentration of interferer molecules increases, their contribution to the noise also increases. While the signal improves with increased receptor concentration, the noise introduced by the interferers becomes more pronounced, which reduces \( \text{SNR}_2 \) to a greater extent. This emphasizes the challenge of balancing receptor concentration to enhance signal detection while managing the interference noise from multiple interferer molecules.

In Fig. \ref{snr_multi}(b), the effect of the interferer binding rate constant \( k_2^+ \) on the SNR is shown for the case with four interferer molecules. As \( k_2^+ \) increases, \( \text{SNR}_1 \) increases slightly and then saturates after a certain point. The variation in \( \text{SNR}_2 \) becomes nearly constant, indicating the presence of a saturation effect. However, the decline from \( \text{SNR}_1 \) to \( \text{SNR}_2 \) is more pronounced compared to the case in Fig. \ref{snr1}(e), where only a single interferer is present. In the multi-interferer scenario, the increase in binding rate of interferers leads to a more significant increase in noise, resulting in a sharper reduction in \( \text{SNR}_2 \).

In Fig. \ref{snr_multi}(c), the effect of interferer molecule concentration on the SNR is shown for the case with multiple interferer molecules. The trends are similar to those observed in Fig. \ref{snr1}(a) with a single interferer. However, the decline from \( \text{SNR}_1 \) to \( \text{SNR}_2 \) is more pronounced, as expected, due to the cumulative noise contribution from multiple interferers.

\subsection{SEP Analysis}

The Symbol Error Probability (SEP) is analyzed using Weight Shift Keying (WSK) \cite{aktas2022weight}, where 1-bit and 2-bit WSK are implemented with two and four distinct molecules, respectively. SEP is evaluated through ML detection, which is applied by the receiver to determine the decision rule. The decision rule is established using the Gaussian approximation of additive noise, as described in \cite{kuscu2016modeling}. The SEP is calculated by considering the mean output current \( \mu_{I_m} \) and the variance of the output current \( \sigma_{I_m}^2 \) for each symbol \( m \), where \( m = 0, \dots, M-1 \), and the expression for the SEP is given in \cite{kuscu2016modeling} as:

\begin{equation}\label{eq:32}
\begin{split}
 P_e = \frac{1}{2M}\Bigg[\operatorname{erfc} \biggl(\frac{\lambda_1 - \mu_{I_0}}{\sigma_{I_0}\sqrt{2}}\biggl) + \operatorname{erfc} \biggl(\frac{ \mu_{I_{M-1}}-\lambda_{M-1}}{\sigma_{I_{M-1}}\sqrt{2}}\biggl)\biggl) \\
 + \sum_{m=1}^{M-2}\biggl(\operatorname{erfc} \biggl(\frac{\mu_{I_m}-\lambda_m }{\sigma_{I_m}\sqrt{2}}\biggl) + \operatorname{erfc} \biggl(\frac{ \lambda_{m+1} - \mu_{I_{m}}}{\sigma_{I_{m}}\sqrt{2}}\biggl)\Bigg],
\end{split}
\end{equation}

where \( \operatorname{erfc} (x) = \frac{2}{\sqrt{\pi}}\int_{x}^{\infty} e^{-y^2} \,dy \) is the complementary error function, and \( \lambda_m \) represents the decision threshold obtained from the ML rule. 

In this analysis, two distinct formulations of SEP are considered: For \( \text{SEP}_1 \), the mean output current \( \mu_{I_m} \) represents the total contribution from both the target and interferer molecules, expressed as \( \mu_{I_m} = \mu_{I_{\text{sum}}, m} \), while \( \sigma_{I_m}^2 \) is the intrinsic noise variance. For \( \text{SEP}_2 \), the mean output current \( \mu_{I_m} \) corresponds to the contribution from the target molecules only, i.e., \( \mu_{I_m} = \mu_{I_T,m} \), while the variance \( \sigma_{I_m}^2 \) incorporates the noise contribution from interferers, expressed as \( \sigma_{I_m}^2 = \sigma_{I_m}^2 + \mu_{I_I,m}^2 \), where \( \mu_{I_I,m} \) denotes the noise contribution from the interferer molecules. The distinction between \( \text{SEP}_1 \) and \( \text{SEP}_2 \) lies in how the noise from the interferers is considered. \( \text{SEP}_1 \) includes the total noise from both target and interferer molecules but does not separate their individual effects. \( \text{SEP}_2 \), on the other hand, accounts for the specific contributions of the target and interferer molecules to both the mean and variance of the output current, providing a more accurate representation of the system’s performance under real-world conditions.

\begin{figure*}[!t]

\centering
\subfloat[]{\includegraphics[width=0.6\columnwidth]{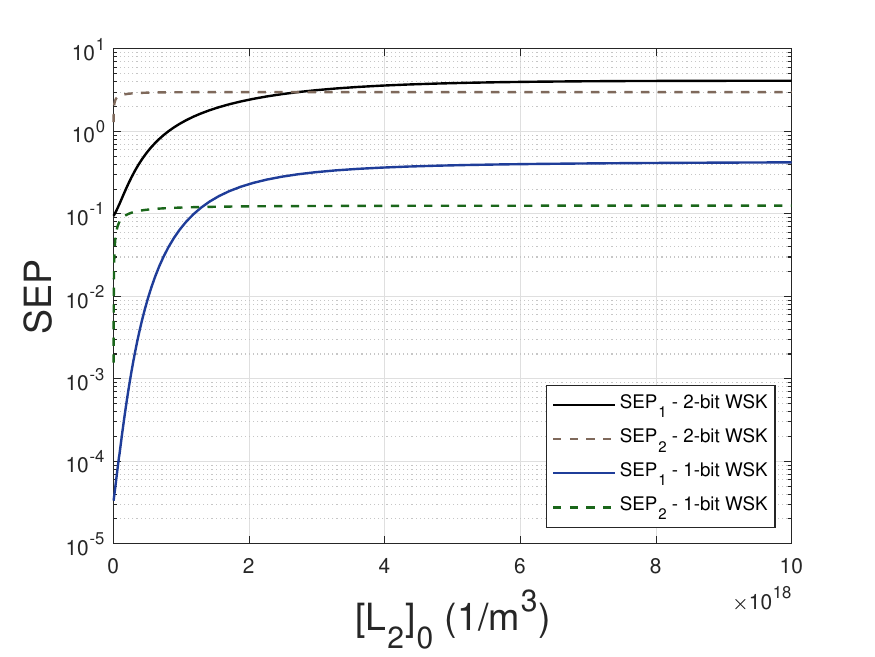}%
\label{fig_first_case1}}
\hfil
\subfloat[]{\includegraphics[width=0.6\columnwidth]{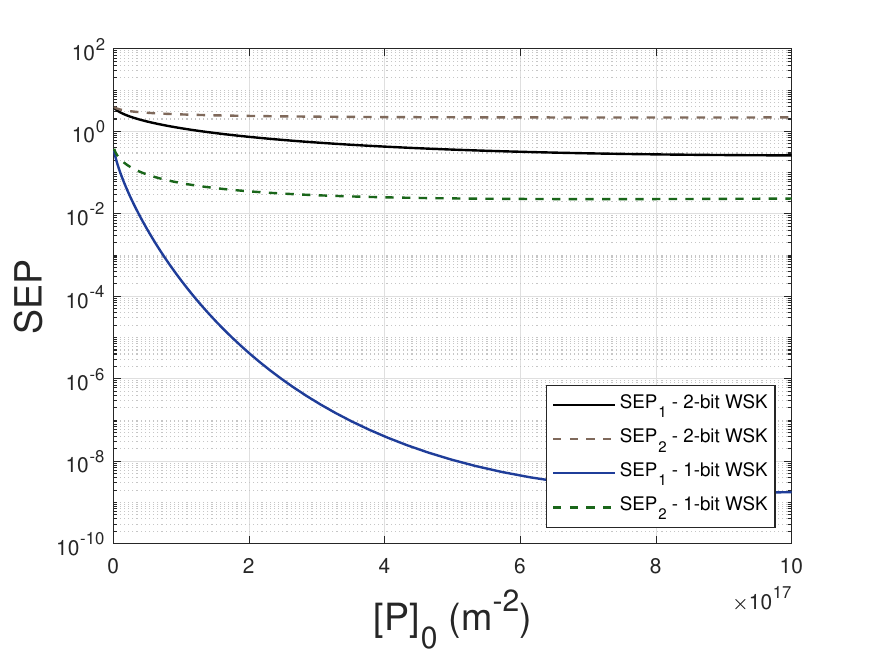}%
\label{fig_second_case2}}
\hfil
\subfloat[]{\includegraphics[width=0.6\columnwidth]{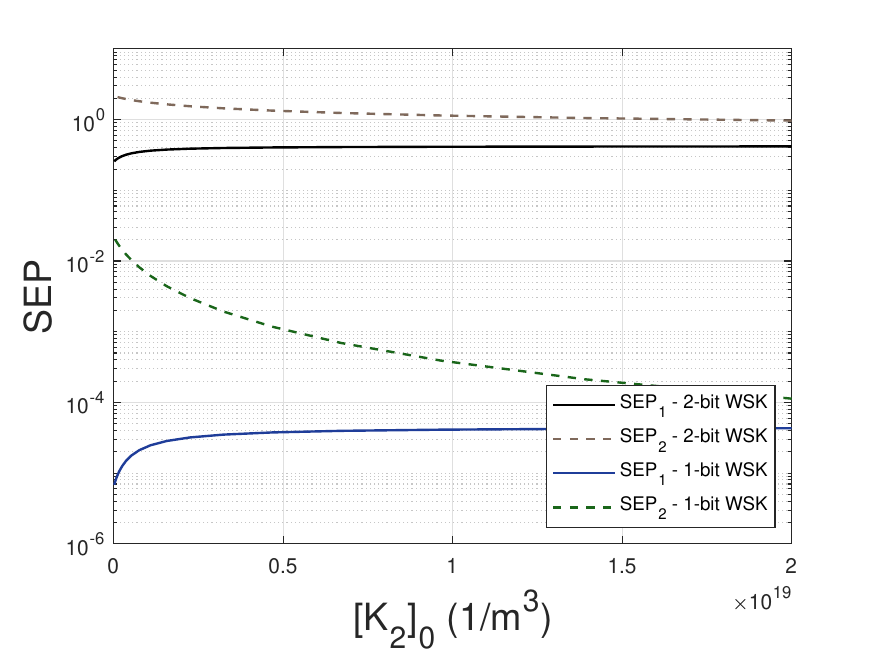}%
\label{fig_third_case3}}
\hfil
\caption{SEP as a function of (a) interferer molecule concentration \( [L_2]_0 \), (b) surface receptor concentration \( [P_0] \), and (c) dissociation constant of interferer molecules \( K_2 \), for both 1-bit and 2-bit WSK.}

\label{sep}
\end{figure*}

In Fig. \ref{sep}(a), the effect of interferer molecule concentration \( [L_2]_0 \) on the SEP is shown for 1-bit WSK and 2-bit WSK. For 1-bit WSK, \( \text{SEP}_1 \) increases as \( [L_2]_0 \) rises, indicating that the system’s performance degrades due to increasing interference from interferer molecules. Since \( \text{SEP}_1 \) does not account for interferers as part of the noise, the error rate grows as the interferer concentration increases. In contrast, \( \text{SEP}_2 \) remains almost flat, suggesting that once interferer noise is considered, the system stabilizes despite increasing concentrations. For 2-bit WSK, the SEP values are higher than those for 1-bit WSK, as expected, because more symbols increase the system’s capacity but also introduce a trade-off by increasing the error rate. The system’s ability to distinguish between signals improves, but the added complexity and interference cause a larger SEP compared to 1-bit WSK. This behavior reflects the inherent trade-off in increasing symbol capacity: while more symbols allow the system to encode more information, they also make the system more susceptible to errors.

In Fig.\ref{sep}(b), the effect of increasing surface receptor concentration \( [P_0] \) on SEP is shown. As expected, higher receptor concentration improves SEP performance for both 1-bit and 2-bit WSK due to enhanced signal detection. Similar to Fig. 10(a), 1-bit WSK outperforms 2-bit WSK, resulting in lower SEP values. Additionally, the SEP\(_2\) curves remain nearly flat, indicating that when the interferer noise is incorporated in the model, the SEP becomes less sensitive to variations in \( [P_0] \).

In Fig. \ref{sep}(c), the effect of the dissociation constant \( K_2 \) for interferer molecules on SEP is shown. Unlike Fig. \ref{sep}(a) and Fig. \ref{sep}(b), where \( \text{SEP}_2 \) remained relatively flat, here \( \text{SEP}_2 \) (green) also decreases with increasing \( K_2 \), but not as sharply as \( \text{SEP}_1 \). This behavior suggests that the dissociation of interferers makes the system more stable, reducing their contribution to the noise and thus improving the system’s performance.

Increasing the dissociation constant \( K_2 \) means that interferer molecules dissociate more readily or quickly. This leads to a faster release of the interferer molecules from their binding sites, reducing the time they spend interacting with the receptors. Since \( \text{SEP}_2 \) incorporates the noise contribution from interferers, the more rapidly the interferers dissociate, the less their noise contributes to the overall system noise. As a result, the system's ability to detect the target signal improves, leading to a lower \( \text{SEP}_2 \) value. Essentially, the faster the interferers dissociate, the less they impact the signal detection, leading to enhanced system performance.

The differences observed between 1-bit WSK and 2-bit WSK are consistent with previous trends: 2-bit WSK still shows higher \( \text{SEP} \) values due to the added complexity and error rate trade-off, but the performance improvement with increasing \( K_2 \) is more noticeable for 1-bit WSK.

\section{Conclusion}

In this paper, we introduced a competitive binding model integrated into the Flexure-FET MC receiver to address the challenges of interference-rich environments. By accounting for the competition between multiple molecular species for receptor binding sites, the framework significantly enhances the receiver's ability to operate in real-world MC systems, where multiple species coexist and interfere with each other. This integration allows for a more accurate representation of receptor occupancy and its dynamics under varying interference conditions, enabling the Flexure-FET MC receiver to operate more efficiently in complex molecular environments.

Our performance analysis demonstrated that this integration improves the reliability and accuracy of the receiver, emphasizing the critical role of managing interferer molecule concentrations in ensuring optimal system performance. The results revealed that, while SNR and SEP are impacted by the presence of interferer molecules, the competitive binding model provides the ability to fine-tune receptor responses, thus optimizing Flexure-FET MC receiver performance in interference-dominated environments. This demonstrates that competitive binding plays a pivotal role in improving the system's sensitivity and robustness to noise and interference.

Additionally, the findings from this work open up new possibilities for future applications in OMC systems. By incorporating competitive binding dynamics, this model can be extended to develop an OMC receiver, enabling non-invasive sensing and real-time detection in fields like healthcare, environmental monitoring, and agriculture. The ability to detect specific odor molecules in complex mixtures offers new opportunities for diagnostics, air quality monitoring, and agriculture. The competitive binding framework enhances molecular communication systems and lays the foundation for future advancements in OMC, driving innovation in environmental sensing and sensory applications.

\bibliographystyle{IEEEtran}
\bibliography{main}

\vspace{11pt}

\begin{IEEEbiography}[{\includegraphics[width=1in,height=1.25in,clip,keepaspectratio]{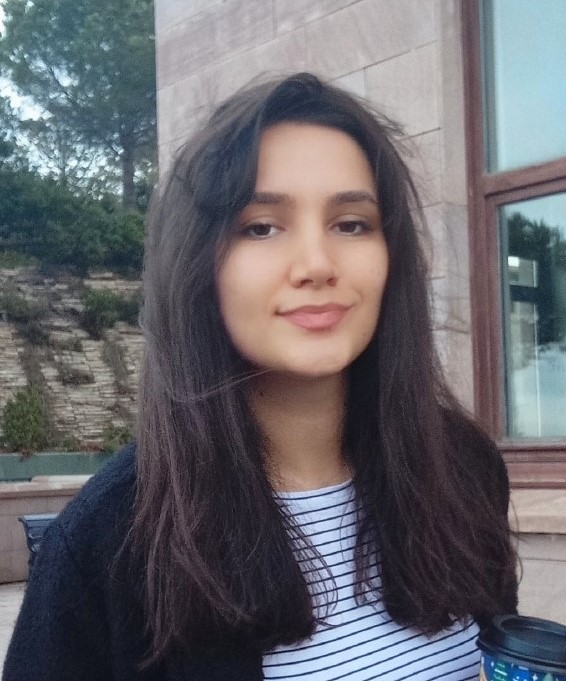}}]{Dilara Aktas}
(Student Member, IEEE) received her BS degree in Electronics and Communication Engineering from Istanbul Technical University in June 2019. She is currently a Research Assistant with the Center for neXt-generation communications at Koç University, Istanbul, Turkey. Her research interests include molecular communications and the Internet of Everything.
\end{IEEEbiography}

\vspace{11pt}

\begin{IEEEbiography}[{\includegraphics[width=1in,height=1.25in,clip,keepaspectratio]{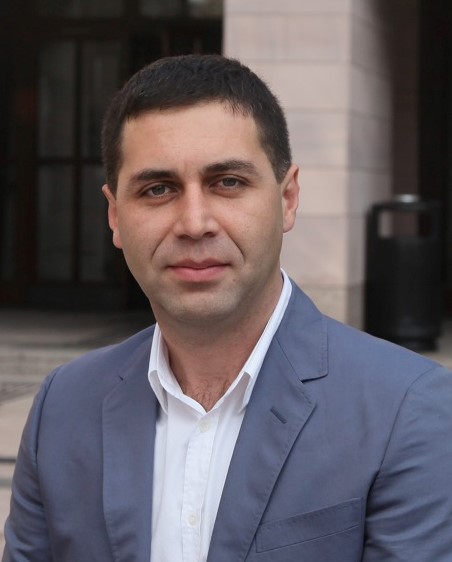}}]{Ozgur B. Akan}
(Fellow, IEEE) received his Ph.D. degree from the Broadband and Wireless Networking Laboratory, School of Electrical and Computer Engineering, Georgia Institute of Technology, Atlanta, in 2004.
He is currently the Head of the Internet of Everything (IoE) Group, Department of Engineering, University of Cambridge, and the Director of the Center for neXt-generation communications at Koç University. His research interests include wireless, nano, molecular communications, and the Internet of Everything.
\end{IEEEbiography}

\vfill

\end{document}